# Nanoscale mapping and spectroscopy of non-radiative hyperbolic modes in hexagonal boron nitride nanostructures

Lisa V. Brown,[¥,1,2] Marcelo Davanco,[¥,1] Zhiyuan Sun,[3] Andrey Kretinin,[4] Yiguo Chen,[5,6] Joseph R. Matson,[7] Igor Vurgaftman,[8] Nicholas Sharac,[9] Alexander Giles,[8] Michael M. Fogler,[3] Takashi Taniguchi,[10] Kenji Watanabe,[10] Kostya Novoselov,[4] Stefan A. Maier,[5,11] Andrea Centrone,[1*] Joshua D. Caldwell[7,8 *]

[1] Center for Nanoscale Science and Technology, National Institute of Standards and Technology, 100 Bureau Dr., Gaithersburg, MD 20899 USA
[2] Maryland Nanocenter, University of Maryland, College Park, MD 20742
[3] Dept. Physics, University of California San Diego 9500 Gilman Dr, La Jolla, CA 92093 USA
[4] School of Physics and Astronomy, University of Manchester, Oxford Rd, Manchester M13 9PL, UK
[5] The Blackett Laboratory, Imperial College London, London SW7 2AZ, UK
[6] Dept. of Electrical and Computer Engineering, National University of Singapore, Singapore 117576
[7] Department of Mechanical Engineering, Vanderbilt University, 101 Olin Hall, Nashville, TN 37212 USA
[8] US Naval Research Laboratory, 4555 Overlook Ave S.W., Washington, DC 20375 USA
[9] NRC Postdoctoral Fellow (Residing at NRL, Washington, DC)
[10] National Institute for Materials Science, 1-1 Maniki, Tsukuba, Ibaraki 305-0044 Japan
[11] Fakultät für Physik, Ludwigs-Maximilians-Universität München, 80799 München, Germany
[¥] Equal contributions to this work.

* josh.caldwell@vanderbilt.edu

* andrea.centrone@nist.gov

**Abstract (150-250 words)**

The inherent crystal anisotropy of hexagonal boron nitride (hBN) sustains naturally hyperbolic phonon polaritons, i.e. polaritons that can propagate with very large wavevectors within the material volume, thereby enabling optical confinement to exceedingly small dimensions. Indeed, previous research has shown that nanometer-scale truncated nanocone hBN cavities, with deep subwavelength dimensions, support three-dimensionally confined optical modes in the mid-infrared. Due to optical selection rules, only a few of many such modes predicted theoretically have been observed experimentally via far-field reflection and scattering-type scanning near-field optical microscopy. The Photothermal induced resonance (PTIR) technique probes optical and vibrational resonances overcoming weak far-field emission by leveraging an atomic force microscope (AFM) probe to transduce local sample expansion due to light absorption. Here we show that PTIR enables the direct observation of previously unobserved, dark hyperbolic modes of hBN nanostructures. Leveraging these optical modes could yield a new degree of control over the electromagnetic near-field concentration, polarization and angular momentum in nanophotonic applications.

Surface phonon polaritons,[1-3] are a low-loss alternative to surface plasmons that enable exciting applications at infrared and terahertz frequencies,[1] including narrow-band thermal emission[2, 4, 5] and yield high quality factors (>300) nanoscale antennas[6-10] and unprecedented modal confinements.[1, 6, 8] These characteristics result in very large Purcell factors that are necessary to achieve strongly enhanced optical processes at the nanoscale, e.g. second harmonic generation in SiC nanopillar arrays.[11] Hexagonal boron nitride (hBN) has recently attracted much interest because of its intrinsic hyperbolicity[7, 12] and the potential for integration with graphene and other van der Waals materials[13-15] to realize novel hybrid architectures.[16, 17] Hyperbolicity is an extreme form of optical anisotropy, in which the values of the real part of the dielectric function along orthogonal crystalline axes are not only different, but in fact opposite in sign. This property enables polaritonic modes with very large wavevectors (i.e. confinement to very small dimensions)[18] and restricts the polariton propagation along directions dictated by the material dielectric function.[19-22] Hexagonal boron nitride is considered the archetype of an ever-broadening library of naturally hyperbolic materials[23-25] and provides an ideal platform for studying hyperbolic phonon-polaritons (HPhPs)[7, 12, 14, 26-29] - deeply sub-diffraction, volume-confined collective optical modes - that enable unique applications in mid-infrared nanophotonics.[21, 22]

In contrast to conventional metals and polar insulators, where polaritons are confined to the material surface, hyperbolic phonon polaritons can freely propagate in a nanostructure volume and must be characterized by three distinct quantum numbers (instead of two for surface polaritons). The choice and physical meaning of these numbers (or modal indices) is geometry-specific. For example, thin hBN slabs are characterized by the in-plane momentum vector $\vec{q} = (q_x, q_y)$, which is a continuous variable, and one branch of index $l = 0, 1, ...$, a discrete number[7, 12, 30] (see Fig S1 of the supplemental information for a slab of with thickness $t$ =256 nm). For nanostructures that confine polaritons in three dimensions,[7, 26, 30] all their modal indices must be discrete. For example, the modes of ideal spheroidal nanoparticles are characterized by the azimuthal angular momentum $m$, the orbital angular momentum $l$, and a radial index $n$. Far-field spectroscopic experiments on three-dimensionally confined truncated nanocones, the so-called 'frustum' shaped nanostructures,[7, 26] have detected a few low-order HPhP resonances ( $l \leq 7, m = 1, n = 0; l \leq 3, m = 0, n = 0$) in two distinct spectral bands. These resonances are qualitatively described by the calculations for ideal spheroidal particles when parametrized with respect to the frustum aspect ratio

$$A_r = d/t, \qquad (1)$$

where $d$ is the average diameter at half-height and $t$ is the film thickness.

The higher-$l$ modes have many desirable properties. They are characterized by a shorter polariton wavelength $\lambda_p = 2\pi/q$ and hence, by a stronger field confinement that can be exploited for sub-diffraction imaging and focusing.[31-33] Although many other higher-order HPhP modes (with $m > 1$ and $n > 0$) have been theoretically predicted,[30] they remained elusive because coupling to far-field radiation is either very weak or forbidden by selection rules. These high-order HPhP modes offer more precise tailoring of the evanescent near-field spatial distributions and better matching of the polariton orbital angular momenta to those of nanoscale absorbers or emitters, potentially modifying the selection rules to make weak or forbidden transitions dominant.[34, 35]

Scattering-type scanning-near-field optical microscopy (s-SNOM) utilizes a sharp, metallized tip as a scanning optical antenna that can aid the excitation of such modes in the near-field and has enabled imaging high-$l$ modes in hBN slabs in real space.[27, 31] Here, we show that another near-field technique, photothermal induced resonance (PTIR) enables direct observation of multiple 'limbs' ($m = 0, n > 0, l = 1,2,3, …$) of the hBN hyperbolic dispersion relation in hBN frustum nanostructures. These measurements give the first unequivocal evidence for the existence of the 'dark' higher order modes.[30] In previous far-field and s-SNOM measurements[7, 26] only the ($l = 1,2,3, …$) 'branches' of the $m = 1, n = 0$ limb were detected. PTIR makes the observation of the higher order modes possible because it detects light absorption by mechanically transducing the sample photothermal expansion, thus bypassing the challenge imposed by weak far-field emission that limits established methods. Consequently, these results highlight the PTIR capability to characterize nanophotonic dark-modes of both artificial meta[18, 36-40] and natural[7, 12, 23, 25] hyperbolic media.

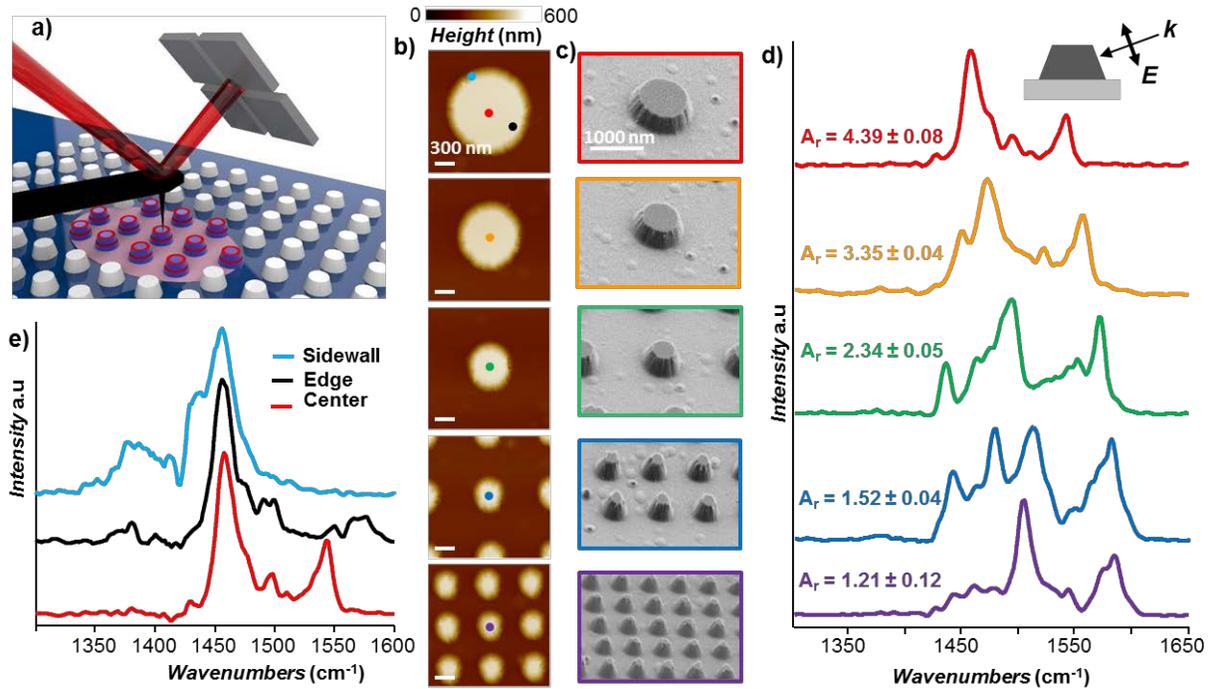

*Fig. 1:* a) Schematic of the PTIR experiment. A pulsed tunable IR laser illuminates the portion of the sample (region highlighted in pink centered around gold-coated AFM probe operating in contact mode). b) AFM topography (scale bars 300 nm) and c) corresponding tilted SEM images (scale bars 1.0 μm) of five representative frustums with different aspect ratio. All the frustums have a thickness of 256 nm ± 4 nm. d) PTIR absorption spectra (p-polarization) obtained by positioning the AFM tip at the center of representative frustums, as indicated in panel b. The schematic of the incident polarization used for the measurements is provided as an inset. e) Color coded position dependent PTIR absorption spectra for the $A_r$ = 4.39 frustum. The spectra in panel d and e are displayed with an offset for clarity.

The PTIR combines the spatial resolution of atomic force microscopy (AFM) with the specificity of absorption spectroscopy. Because PTIR often utilizes infrared (IR) excitation, it is also known as AFM-IR. Similar to s-SNOM, in PTIR a metallized AFM tip is used to overcome the momentum mismatch and/or enhance coupling between the far-field radiation and the deeply sub-diffraction polaritonic modes. In contrast to s-SNOM, where the tip is operated in tapping mode and the scattered light is detected, in the PTIR experiments presented here, the AFM probe is in contact with the sample (contact mode) and acts as a near-field mechanical detector. A detailed comparison between the working principles of PTIR and s-SNOM and corresponding applications is available elsewhere.[41] The PTIR signal transduction typically follows the sequence: optical energy into vibrational energy (absorption), absorbed energy into heat, heat into mechanical thermal expansion, thermal expansion into cantilever motion, and cantilever motion into AFM photodetector signal.[42] Although novel nanoscale probes with integrated cavity opto-mechanics,[42] can track the fast sample thermal expansion dynamics, the conventional cantilevers used here are too slow to capture that process and are kicked into oscillation by the sample expansion similar to a struck tuning fork. Because the amplitude of the PTIR signal (cantilever oscillation) is proportional to the energy absorbed by the sample,[43, 44] PTIR spectra enable mapping of electronic bandgap[45] and chemical composition[46] at the nanoscale, with applications in biology,[47, 48] photovoltaics,[49] polymer science,[50, 51] pharmaceutics[52] and plasmonics,[53-55] as reviewed elsewhere.[41, 56] For example, PTIR recently enabled the observation of 'dark' modes in plasmonic antennas[53, 57] and the imaging of HPhPs modes in micrometer-sized hBN flakes.[58]

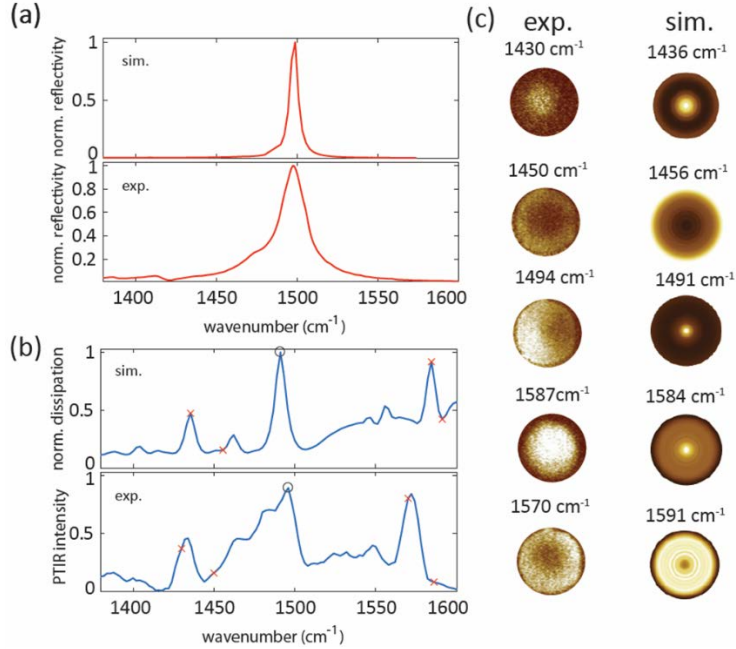

*Fig. 2: Simulated (top) and experimental (bottom) a) far-field reflectance and b) PTIR absorption spectra from the $A_r = 2.34$ frustum. c) PTIR absorption maps for the spectral position labeled (1 to 5) in panel b). d) Corresponding calculated PTIR maps obtained by integrating the power loss over the frustum volume for each position of the point dipole (50 nm above the frustum surface) that simulate the tip position in the PTIR experiments.*

The PTIR set up used in this work consists of a wavelength-tunable, pulsed laser that illuminates a portion of the sample (≈ 50 μm diameter centered around the AFM probe) with an angle of ≈ 20° with respect to the sample surface (Fig. 1a). The gold-coated Si AFM probe provides the momentum-matching necessary for efficiently coupling the incident light into the highly confined, large-wavevector polaritonic modes. Five periodic arrays of hBN frustum-shaped nanostructures with $d$ ranging from 312 nm ± 26 nm to 1120 nm ± 22 nm (as measured by AFM; Fig. 1b), similar to those previously measured by far-field reflectance/transmittance[7] and s-SNOM,[26] are studied here. The uncertainties in the nanostructure dimensions represent one standard deviation in the measurements of seven nominally identical structures. The nanostructures, have a thickness of $t$ = 256 nm ± 4 nm and are characterized by $A_r$ ranging from 1.21 ± 0.12 to 4.39 ± 0.08, where the $A_r$ uncertainty is determined by the uncertainty in the measurement of the frustum diameters. The corresponding scanning electron microscope (SEM) images are provided in Fig. 1c, with details on the sample fabrication available elsewhere.[7, 26] The hBN HPhP modes were observed within the hBN upper Reststrahlen band,[7, 12] which extends between the frequency of the in-plane longitudinal $\omega_{LO} \approx$ 1610 cm$^{-1}$ and transverse optical phonons $\omega_{TO} \approx$ 1360 cm$^{-1}$.[7, 12, 59] Local PTIR absorption spectra (Fig. 1 d,e) were obtained for each cone by placing the AFM tip in specific locations on the frustums and plotting the amplitude of the cantilever-induced oscillations as a function of laser frequency. In addition, PTIR absorption maps were obtained by scanning the AFM tip while illuminating the sample at a given incident frequency and measuring the amplitude of the cantilever-induced oscillation as a function of tip location. PTIR experiments can be obtained with either *s*- (in plane) or *p*-polarized light. Because of the incident light angle, illumination with *p*-polarization provides both an in-plane and an out-of-plane electric field component, the latter being amplified by the larger vertical polarizability of the gold-coated probe. All PTIR data provided in the main text were obtained using p-polarized light (inset of Fig. 1d) since this illumination condition provides the most interesting results. For completeness, PTIR data obtained with s-polarized light are available in the Supplemental Information (Figs. S2 and S3).

The PTIR spectra from the center of representative frustums (Fig. 1d) highlight a broad range of optical modes and a systematic mode dispersion as a function of $A_r$. Although some of these resonances appear to directly correlate to those previously observed in both far-field[7] and s-SNOM[26] measurements, several additional, strong modes are observed. Particularly, many of the strongest resonances are observed at frequencies above ≈ 1550 cm$^{-1}$ (i.e. above the $m = 1, n = 0, l = 1$ resonance for the smallest $A_r$ frustums) and were not reported for any of the arrays in prior works. Because each of the modes observed in the PTIR spectra show a systematic dispersion with the $A_r$, we deduce that these resonances are indeed HPhP resonance modes. Furthermore, in contrast with previous s-SNOM observations,[26] PTIR spectra obtained at different locations on a given frustum (see Fig. 1e) display clear spectral variations. This is most likely related to the how well the AFM tip can launch HPhPs while in contact with the frustum, rather than to the launching of the HPhPs by the sharp frustum corners.

The Fourier-Transform Infrared (FTIR) reflection spectrum for the $A_r$ = 2.34 frustrum (Fig. 2a bottom panel) is dominated by a strong resonance at 1494 cm$^{-1}$ along with two weaker resonances located at 1413 cm$^{-1}$ and 1384 cm$^{-1}$. These modes have been previously identified and assigned as $TM_{ml}^U$ where the subscripts are the modal indices $m = 1$ and $l = 1, 2, 3$, respectively, with the 'U' superscript referring to the upper Reststhralen band. The corresponding resonances were identified at ≈ 1491 cm$^{-1}$, ≈ 1414 cm$^{-1}$ and ≈ 1389 cm$^{-1}$ within the PTIR spectrum (Fig. 2b bottom panel). We note that because the HPhPs in the upper Reststrahlen band with $m$ = 0 and/or radial index $n > 0$ cannot be directly stimulated or measured in the far-field, the radial quantum number $n$ (formerly denoted $r$) was ignored in previous work.[7] Hereafter, all three quantum numbers, $l$, $m$, and $n$ are used to describe the broad range of additional modes observed in the PTIR experiments,. Consequently, the previously observed resonances in hBN nanostructures correspond to the $(lmn) = (l10)$ modes in the notation used here. The assignment and spectral positions for the first three branches of the $(l10)$ limb match well between the experimental PTIR and far-field reflection data (Fig. 3b), as well as with resonances calculated for ellipsoidal nanoparticles (Fig. 3a). Although the structures discussed here are not ellipsoidal, an analytical solution for ellipsoidal particles[30] previously provided a good qualitative description of the optical modes and the trends in the spectral dispersion with respect to changing $A_r$ within frustum-shaped hBN nanoparticles.[7]

The extension of the analytical calculations beyond the $(l10)$ case to include those with $m = 0$ and $n > 0$ in the upper Reststrahlen band, offers a rich and complex dispersion relationship (see Fig. 3a). Here, the hyperbolic dispersion is represented as a function of $A_r$ with the gray-scale corresponding ro the calculated Purcell factor for the mode. To limit the complexity of the plots, only HPhPs with $m$ = 0 or $m$ = 1 and $n = [(l - m)/2] \leq 2$ are given. In this dispersion plot of Fig 3a, we observe a series of limbs each corresponding to different $m$ and/or $n$. Each limb is composed of several branches of varying $l$, with each limb designated by solid lines of the same color in Fig 3a, which are provided as a guide to the eye. The previously reported $(l10)$ limb is highlighted in orange. The previously unreported limbs [$(l02)$; blue], [$(l11)$; red], and [$(l01)$; green] are also included. We now compare these theoretical curves with the observations. Close examination of the PTIR spectra for each $A_r$ demonstrates several series of dispersing resonant modes. Using the dispersion plot in Fig. 3a as a guide, we can employ least-squares fitting of the resonances observed in the PTIR absorption spectra (representative fit results in Fig S5), to assign and group the experimentally observed modes into corresponding limbs and branches (Fig 3b) that qualitatively match the analytical calculations (Fig 3a). While the sheer number of resonant modes observed in the PTIR experiment makes the unambiguous identification of each resonance challenging, the good qualitative agreement supports our assignments. Furthermore, because least-square fitting of the far-field reflectance data enabled the identification of the $(l10)$ (open orange symbols in Fig. 3b) we could compare the HPhP spectral positions of these modes with those observed in the PTIR spectra directly. The good match between the FTIR and PTIR spectral positions (Fig 3b) adds additional confidence to the spectral assignments. Regardless of the specific assignments, our analysis clearly demonstrates that the additional resonances observed here are consistent with those previously anticipated by theory,[30] including the modes occurring at frequencies higher than the frequency of the (110) mode. As noted above, with the exception of the $(l10)$ limb, the hBN resonances have not been experimentally observed previously owing to the limitations imposed by the far-field selection rules, highlighting the power of the PTIR technique for nanophotonic modal observation and analysis.

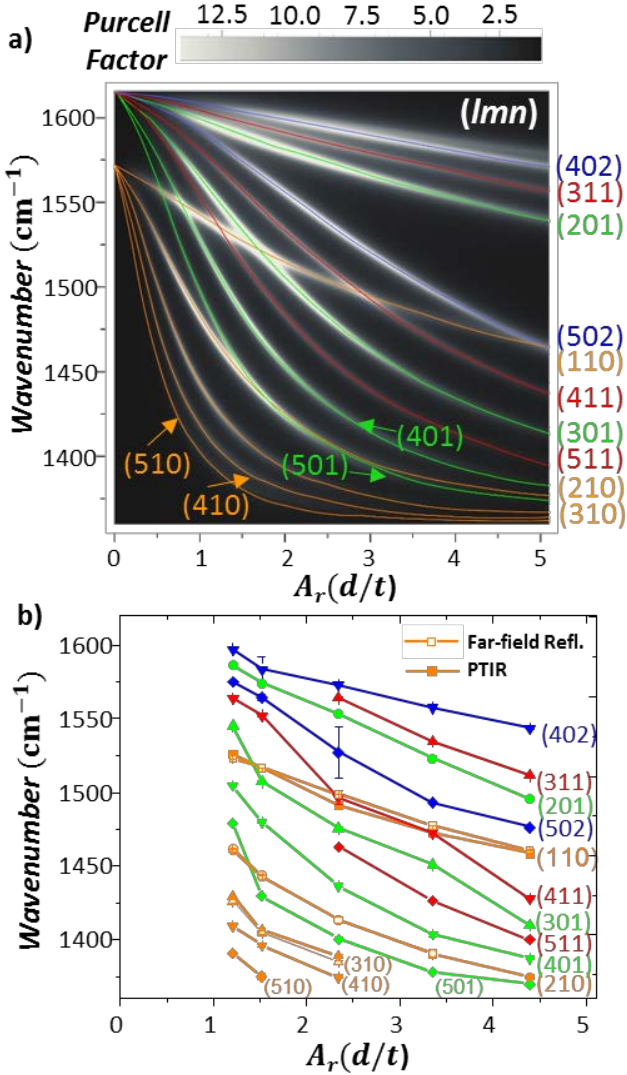

*Fig. 3*: a) Analytical calculations of the HPhP dispersion for hBN ellipsoidal particles as a function of the Ar. Four specific 'limbs' are observed with multiple dispersing polaritonic modes. The gray-scale represents the corresponding Purcell factor, while the colored lines are provided as a guide to the eye and identify the $(l10)$ (orange traces), $(l11)$ (red traces), $(l01)$ (green traces) and $(l02)$ (blue traces) limbs of the $A_r$-dependent HPhP dispersion for hBN ellipsoidal particles. b) PTIR (full symbols) and FTIR (open symbols) experimental $A_r$-dependent dispersion from each of the five frustums measured in Fig. 1b. The modal assignments are in good qualitative agreement with the analytical calculations. The color lines and symbols, along with the (lmn) designations are provided for easier comparison with the calculations. The dispersion of the three observed resonances within the far-field reflection are also provided as the open circles and correspond to the (110), (210) and (310) branches. The error bars represent a single standard deviation of the resonance peak position uncertainties determined by the least-squares fitting algorithm. Some uncertainties are smaller than the dot size.

Electromagnetic simulations based on the finite-element method (FEM) offer further insight into the origin of the polaritonic resonances observed in the PTIR measurement. First, a simulation of plane-wave scattering from a frustum with $A_r$ = 2.34 is performed (see Methods for details) and yields the reflection spectrum (Fig. 2a, top panel) where the prominent peak at 1491 cm$^{-1}$ corresponds to the bright (110) resonance, in good agreement with the FTIR spectrum. Importantly, the additional polariton resonances detected in the PTIR measurement are not observed in the simulated reflection spectrum, reinforcing our postulate that they are due to polaritonic modes inaccessible by far-field excitation. We next performed a series of FEM simulations in which the same frustum was illuminated by a radiating electric dipole located 50 nm above the top surface. As we show next, this procedure emulated the presence of the AFM tip in the PTIR experiment, leading to the excitation of polariton waves through high spatial frequency electromagnetic field components localized at the launching point. From steady-state field solutions to the dipole simulations, we obtained the electromagnetic power loss in the frustum, which we assumed to lead to an isotropic expansion of the geometry, proportional to the PTIR signal (details in the Methods and discussion below). Following this procedure for a dipole located at the center of the frustum, we obtained the dissipation spectrum shown in the top panel of Fig. 2b that is qualitatively well-matched to the PTIR spectrum obtained with the tip at the center of the $A_r$ = 2.34 frustum (Fig. 2b, bottom panel). Specifically, prominent peaks and valleys labeled 1 to 5 are observed at roughly the same frequencies in both the PTIR and calculated spectra, indicating that the radiating dipole excites approximately the same polariton resonances as the PTIR tip.

A series of PTIR absorption maps (Fig. 2c) reveal the evolution of the HPhP near-field distributions for the frustum nanostructures with $A_r$ = 2.34 as a function of frequency. The corresponding PTIR maps for the frustums with different $A_r$ are provided in the Supplemental Information, Fig S6. As reported previously,[7, 26] the polariton near-field patterns exhibit a continuous evolution with respect to frequency, due to the frequency-dependent polariton propagation angle.[18, 26] Consequently, the field profiles that give rise to the PTIR absorption maps are decoupled from the resonances themselves, thereby providing an opportunity to tailor the near-field amplitude and the mode spatial distribution with nanostructure design. However, because the near-field patterns are not unique to a specific resonance they should not be directly labeled as such. The observed patterns are the result strong localized absorption occurring at spatial locations where the ray-like trajectory[30] of the HPhPs launched by the PTIR tip intersects with the exposed surface of the frustum. Although a spatial resolution as high as 20 nm[60] has been reported for contact-mode PTIR experiments, it is also known that the PTIR spatial resolution is somewhat dependent on the sample thermomechanical properties.[41, 61] The PTIR images in this work were recorded with a horizontal pixel resolution between 6 nm and 3.2 nm depending on the scan size (see Methods). However, we estimate the spatial resolution to be of the order of 35 nm, determined from the sharpest features observed in the PTIR maps, i.e. comparable to the tip size. Such high-spatial resolution is somewhat surprising considering the high in-plane thermal conductivity of hBN and the relatively short lifetime of hBN HPhPs (< 10 ps),[27, 29] that is much shorter than the laser pulse length in the PTIR experiments (≈ 500 ns). Given these conditions, one would expect that the sample would thermalize within the duration of the laser pulse, leading to a homogeneous signal with intensity proportional to the sample thickness, as is the case for experiments with *s*-polarization (Supplemental Information), or for modes coupling to the far-field. The heterogeneities in the PTIR maps with *p*-polarization (Fig. 2c and Fig S6) are interpreted to reflect the location-specific coupling efficiency between the free-space wavelength and the HPhPs mediated by the AFM-probe. In other words, given an incident frequency, the gold-coated tip couples light efficiently to the nanostructure only at locations where the corresponding HPhP near-fields are strong. The hBN nanostructure still thermalizes within the 500 ns long laser pulse; however, the amount of absorbed energy (PTIR intensity) is proportional to the local light-matter coupling strength.

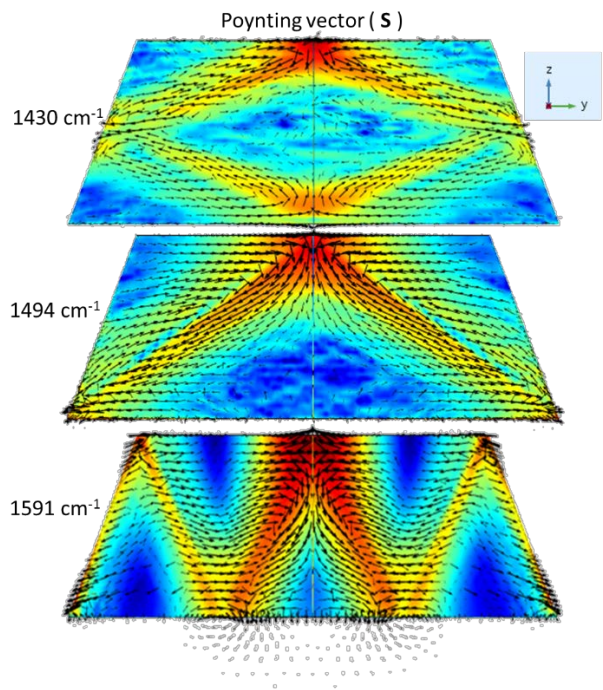

**Fig. 4**: Three dimensional electromagnetic calculations of the cross sectional electromagnetic field amplitude for a frustum with $A_r$ =2.3 illuminated by an electric point dipole located the center of the frustum, 50 nm above the top surface, at 1436 cm$^{-1}$ (top), 1491 cm$^{-1}$ (middle) and 1591 cm$^{-1}$ (bottom). The overlaid arrows designate the Poynting vector (**S**) for the polaritonic modes. Note that the direction of the field cycling and thus the predominant direction of the Poynting vector are reversed between frequencies lower (1430 cm-1) and above (1591 cm-1) the principle (110) resonance of the hBN frustum.

Dipole-excitation, electromagnetic simulations using dipole-excitation were also employed in an attempt to replicate and understand these features in the PTIR absorption maps. For this purpose, the dipole excitation source, at selected wavenumbers, was scanned across the frustum top surface. At each location, the total electromagnetic power loss in the nanostructure was calculated (details in the Methods). The calculations are restricted to the frustum top surface because they could not easily provide information for the nanostructure sidewalls. The calculated electromagnetic power loss maps in Fig. 2d are qualitatively similar to the corresponding PTIR images in Fig. 2c (see Fig. S6 for the PTIR images that include the frustum sidewall) and exhibit a gradually changing absorption profile. A distinguishing feature of both the experimental and calculated data sequence is the frequency-dependent evolution of the absorption distribution. It alternates between featuring a maximum near the center of the frustum ('1', '3', '4' in Fig. 2c) and a donut-shape with a minimum at the center ('2' and '4' in Fig. 2c). These maps, along with the dependence of the PTIR spectra with probe position (Fig. 1e) imply that the AFM probe plays a significant role in the launching of the polaritons. This is in direct contrast to s-SNOM measurements where the presence of the tip induces a minor perturbation to the dielectric environment.[26] Calculations of the cross-sectional electric field, together with the corresponding power flux profiles obtained with the dipole source fixed 50 nm above the center of the frustum top-surface highlight trajectories of the polaritonic rays as a function of the frequency (Fig 4). As demonstrated by Caldwell *et al.*[7] and Giles *et al.*,[29] the frequency dependent propagation angle ($\theta$) with respect to the *z*-axis within these frustums is dictated by the hyperbolic polariton propagation condition:

$$\theta(\omega) = \arctan\left(\frac{1}{i}\frac{\sqrt{\varepsilon_t(\omega)}}{\sqrt{\varepsilon_z(\omega)}}\right), \quad (2)$$

where the subscripts $t$ and $z$ designate, respectively, the in-plane and the out-of-plane axes of the crystal; $\varepsilon_j$ is the frequency-dependent dielectric function (permittivity) of the material along the $j$-axis. At the principal (110) resonance frequency (1494 cm$^{-1}$, Fig 4b), it is apparent that the power flux from the dipole emitter propagates at the anomalous angle $\theta$ to the lower corner of the frustum, then returns back to the center of the top surface. Below 1494 cm$^{-1}$ (i.e. at 1430 cm$^{-1}$, Fig. 4a), a similar power flow pattern is observed, except that the larger propagation angle leads to the HPhP reflection at the frustum sidewall. These types of HPhP trajectories were described previously by Caldwell *et al.*[7] and observed experimentally in higher aspect-ratio frustums.[29] However, above the 1494 cm$^{-1}$ principal mode, however, (e.g., at 1591cm$^{-1}$, Fig. 4c) the power flows at a smaller angle $\theta$, leading to polariton reflections against the frustum's *top* and *bottom* surfaces. This effect has not been observed previously and is characterized by a Poynting vector predominantly oriented in the transverse direction, in contrast with the longitudinal ($z$-axis) direction of the previous cases. This wide variation in the power-flow direction could be exploited in hBN applications; for example, it could impact the coupling between the HPhPs and local emitters placed next to the hBN nanostructures. We note that the resonances shown in Fig. 4 are supported in frustums of different $A_r$, albeit with a spectrally shifted frequencies (Fig. S7 of the SI that shows the cross-sectional electric-field magnitude and Poynting vector plots for resonances of frustums with four different aspect ratios).

In summary, the PTIR technique was used here to provide the first observation of non-radiative HPhPs modes in hBN, enabling the detection of up to five branches of four different limbs of the $A_r$-dependent hyperbolic dispersion relationship. Such non-radiative modes were predicted theoretically,[30] however, their far-field selection rules have precluded their direct observation in previous experiments. The PTIR near-field detection, mediated by the sample thermal expansion, enables the detection of these dark HPhP modes. The comparison of the rich PTIR spectra with the calculations of the $A_r$-dependent dispersion for ellipsoidal nanoparticles, for which an analytical solution is available, provided the basis for the assignment of the quantum numbers defining each limb and branch to the experimentally observed modes. The PTIR absorption maps also enabled the visualization of the frequency-dependent evolution of the local resonant absorption that closely mirrors the trends observed in electromagnetic near-field simulations. Such simulations indicate that the power-flow trajectories of confined HPhP rays are a strong function of frequency. This suggests that the local near-field emission profile and/or the propagation properties of a hyperbolic waveguide can be directly modified through local nanostructure design in a manner that is decoupled from control of the individual polariton frequencies. More broadly, these results also highlight the capabilities of the PTIR method in characterizing nanophotonic materials and nanostructures that are complementary to the more traditional techniques such as far-field reflection/transmission spectroscopy and s-SNOM.

**Acknowledgements**

L.V.B. acknowledges support under the Cooperative Research Agreement between the University of Maryland and the National Institute of Standards and Technology Center for Nanoscale Science and Technology, Award 70NANB14H209, through the University of Maryland. J.D.C., I.V. and A.J.G. were funded by the Office of Naval Research (ONR) with the funds distributed by the Nanoscience Institute at the Naval Research Laboratory.  N.S. acknowledges the support of the National Research Council (NRC)/NRL Postdoctoral Fellowship Program. A.V.K. and K.S.N. acknowledge support from the EPSRC (U.K.), the Royal Society (U.K.), European Research Council, and the EC-FET European Graphene Flagship. M.M.F. and Z.S. were supported by the Office of Naval Research under Grant N00014-15-1-2671, by the National Science Foundation under Grant No. ECCS-1640173, and by the Semiconductor Research Corporation (SRC) through the SRC-NRI Center for Excitonic Devices, research 2701.002. S.A.M. acknowledges ONR Global, the EPSRC Reactive Plasmonics Programme EP/M013812/1, and the Lee-Lucas Chair in Physics

**Methods**

Sample Fabrication

The hBN crystals were grown using a high-pressure/high-temperature method[62, 63] and were exfoliated and transferred onto a quartz substrate. The flake thickness was initially measured via atomic force microscopy and measured again after the fabrication of the frustums. The sample was coated with a bilayer of PMMA and patterned with electron beam lithography using an Al hard mask created via electron-beam evaporation and liftoff. The hBN nanostructures were fabricated via reactive ion etching in an oxygen environment. The remaining metal was then removed with wet chemical etchants. Further details of the fabrication process can be found in the literature.[7]

PTIR measurements

PTIR spectra and images were obtained with a commercial PTIR instrument interfaced with a quantum cascade laser tunable from 1934 cm$^{-1}$ (5.17 μm) to 1126 cm$^{-1}$ (8.88 μm) that illuminates the sample at ≈ 20° from the sample plane. The laser repetition rate was set to 1 kHz and the pulse length to 500 ns (0.05 % duty cycle); the laser spot size at the sample was ≈ 50 μm in diameter. PTIR spectra were obtained by averaging the cantilever deflection amplitude from 512 individual laser pulses at each wavelength and tuning the laser at intervals of 2 cm$^{-1}$. Up to four spectra were acquired and averaged for each tip location, and smoothed by considering 7 adjacent points (see Fig. S7 for the comparison between smoothed and unsmoothed data). AFM topography and PTIR maps were acquired simultaneously illuminating the sample at constant wavelength, by synchronizing the signal acquisition so that the PTIR signal in each pixel was an average over 32 laser pulses.[43] All PTIR and topography images were obtained with a 0.06 Hz scan rate over a areas ranging from 0.8 μm x 0.8 μm to 1.5 μm x 1.5 μm depending on the nanostructure size (i.e. probe velocity between 50 nm/s and 100 nm/s). The images are composed of 250 pixels in the horizontal direction and 100 pixels in the vertical direction, that yield a horizontal pixel resolution of 6 nm, 5 nm, 4 nm , and 3.2 nm for the frustum with $A_r$ =4.39, 3.35, 2.34, and 1.52 , respectively. All PTIR experiments were obtained using p-polarization unless otherwise noted. Commercially available 450 μm long gold-coated silicon contact-mode AFM probes with a nominal spring constant between 0.07 N/m and 0.4 N/m were used for all the PTIR experiments.

Electromagnetic Simulations

*Theoretical FTIR Reflection Spectrum*. The theoretical FTIR reflection spectrum shown in the top panel of Fig. 2(a) was obtained with a finite-element based simulation of a plane-wave scattering against a hBN truncated cone (frustum) with height 265 nm and bottom and top diameter of 510 nm and 710 nm respectively ($A_r$ = 2.3), sitting on top of a substrate with refractive index 1.2069 (i.e. Quartz). The complex permittivity of the hBN frustum was the same as in ref. 11. A scattered-wave formulation was adopted, in which the solution of a *p*-polarized plane wave scattering against the planar air-quartz interface at a 20° angle from the sample surface was used to excite the frustum. The intensity of the frustum-scattered field, normalized to the incoming plane wave intensity, was determined to yield the simulated FTIR reflection spectrum.

*Theoretical PTIR Spectrum.* The theoretical PTIR spectrum (Fig. 2b) was calculated via a finite-element method simulation of an electric point dipole radiating 50 nm above the center of the frustum, and oriented normally to its top surface. Perfectly-matched layers (PMLs) surrounded by the frustum in order to emulate open domains in all directions. The steady-state electric field (**E**) inside the frustum, and the electromagnetic power loss per unit volume ($Q_h = 2\pi f \, \varepsilon_0 \varepsilon''|\mathbf{E}|^2$) were obtained in the spectral range between 1380 cm$^{-1}$ and 1600 cm$^{-1}$ (here $f$ is the frequency, $\varepsilon_0$ is the vacuum permittivity and $\varepsilon''$ is the imaginary part of the complex hBN permittivity). The normalized calculated PTIR spectrum in Fig. 2(b) was obtained integrating $Q_h$ over the volume of the entire frustum based on the assumption that the expansion of the frustum is isotropic and proportional to $Q_h$. The choice of 50 nm dipole distance from the frustum surface is justified by the fact that it maximized the visibility of the prominent features in simulation in most cases.

*Theoretical PTIR Images.* The theoretical PTIR maps (Fig. 2c) were obtained by calculating the total electromagnetic power loss $Q_h$ in the hBN cone (as discussed above), at selected wavenumbers for dipole positions varying across the entire frustum top surface.

Analytical Calculations

The hBN nanoparticle was modeled by a spheroid. Its polaritonic eigenmodes were found solving numerically the transcendental equation given in previous work. (Note the small difference in notations: the mode indices $l, m, n$ here correspond to $l, m, r$ in Refs. [7] and [30]). We represent the AFM tip by a point dipole emitter placed above the north pole of the spheroid. The dipole axis is specified by a unit vector **d**, which we chose to be 45 degrees with respect to the equatorial plane of the spheroid plane of the spheroid. We assume that the PTIR signal is proportional to the Purcell factor of the emitter, $f = 1 + (3/2)(\lambda_0/2\pi)^3 \text{Im}(\mathbf{d} \cdot \mathbf{E})$, where $\lambda_0$ is the vacuum wavelength of light and **E** is the reflected electric field at the dipole's location. The calculation of the Purcell factor is performed by expanding **E** in spheroidal harmonics according to the formulas given in the Supplementary Material of Ref. [30].

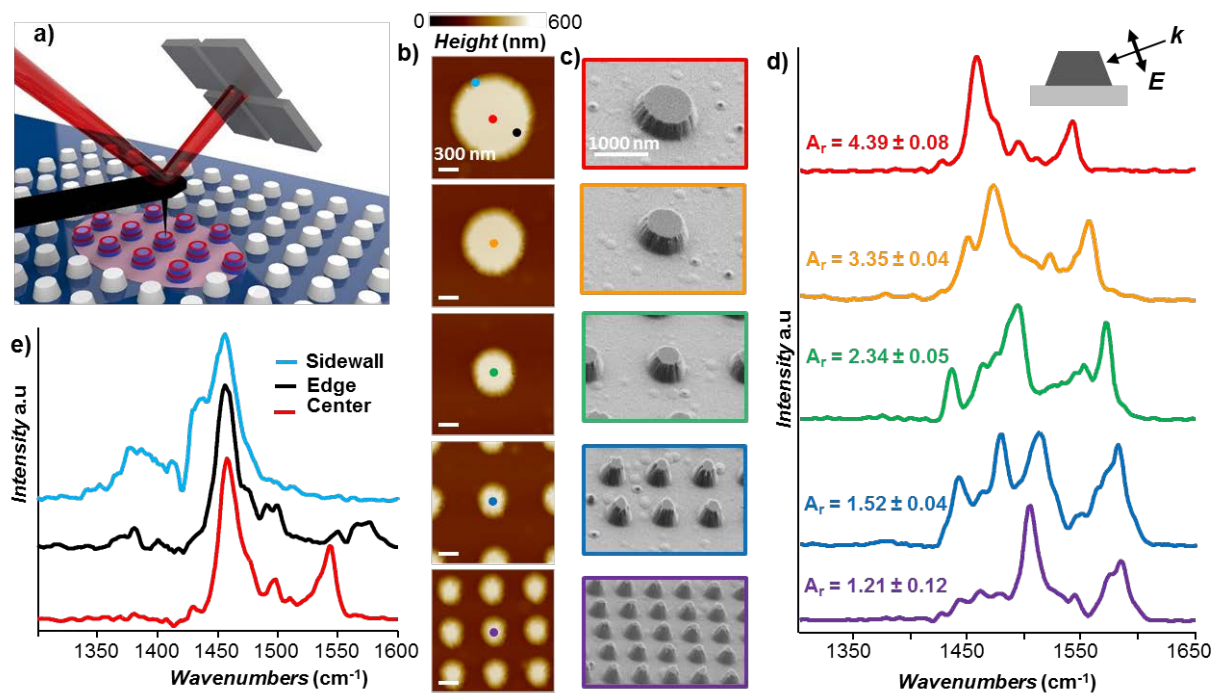

**Fig. 1:** a) Schematic of the PTIR experiment. A pulsed tunable IR laser illuminates the portion of the sample (region highlighted in pink centered around gold-coated AFM probe operating in contact mode). b) AFM topography (scale bars 300 nm) and c) corresponding tilted SEM images (scale bars 1.0 µm) of five representative frustums with different aspect ratio. All the frustums have a thickness of 256 nm ± 4 nm. d) PTIR absorption spectra (p-polarization) obtained by positioning the AFM tip at the center of representative frustums, as indicated in panel b. The schematic of the incident polarization used for the measurements is provided as an inset. e) Color coded position dependent PTIR absorption spectra for the $A_r$ = 4.39 frustum. The spectra in panel d and e are displayed with an offset for clarity.

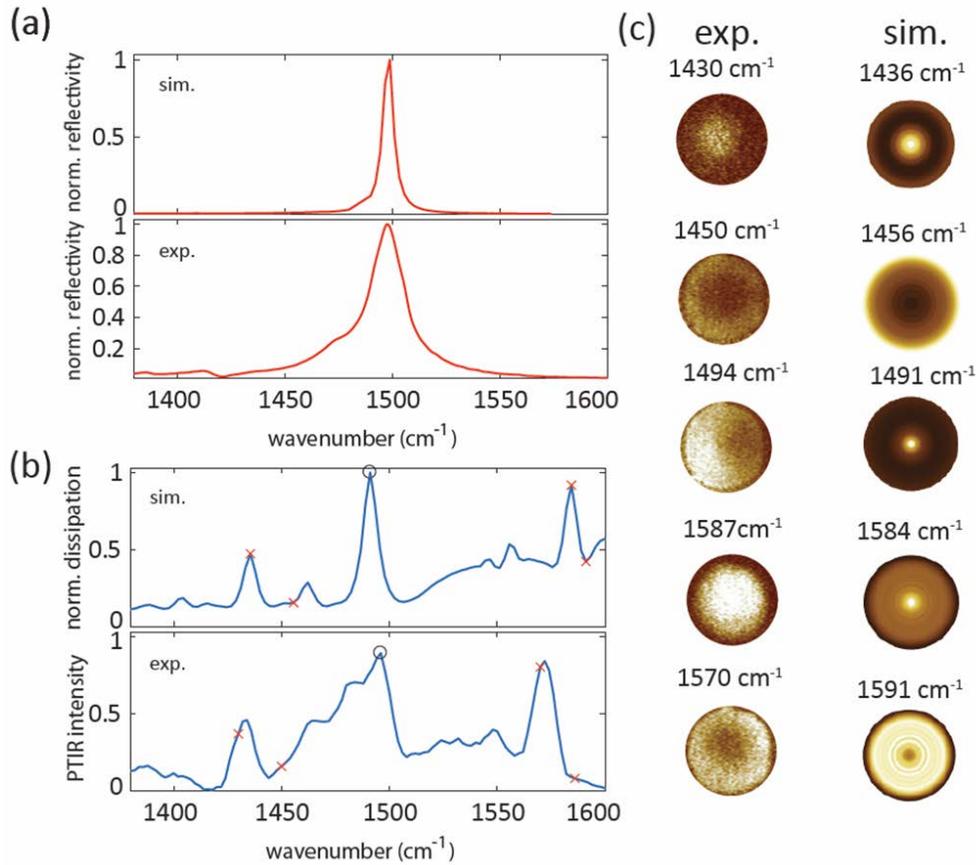

**Fig. 2:** Simulated (top) and experimental (bottom) a) far-field reflectance and b) PTIR absorption spectra from the $A_r = 2.34$ frustum. c) PTIR absorption maps for the spectral position labeled (1 to 5) in panel b). d) corresponding calculated PTIR maps obtained by integrating the power loss over the frustum volume for each position of the point dipole (50 nm above the frustum surface) that simulate the tip position in the PTIR experiments

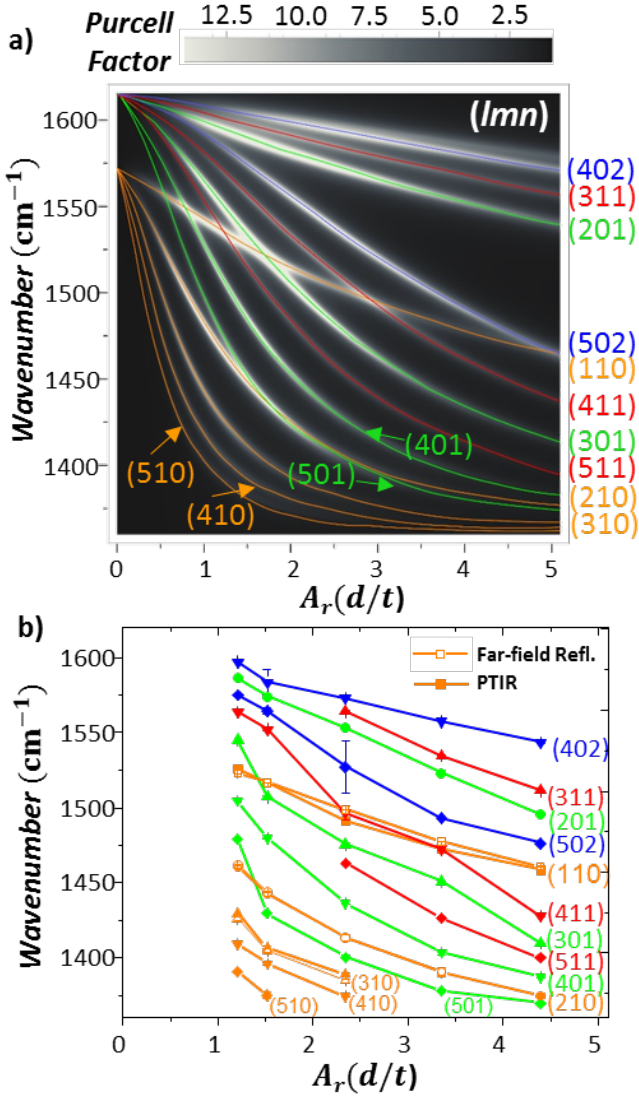

**Fig. 3**: a) Analytical calculations of the HPhP dispersion for hBN ellipsoidal particles as a function of the Ar. Four specific 'limbs' are observed with multiple dispersing polaritonic modes. The gray-scale represents the corresponding Purcell factor, while the colored lines are provided as a guide to the eye and identify the $(l10)$ (orange traces), $(l11)$ (red traces), $(l01)$ (green traces) and $(l02)$ (blue traces) limbs of the $A_r$-dependent HPhP dispersion for hBN ellipsoidal particles. b) PTIR (full symbols) and FTIR (open symbols) experimental $A_r$-dependent dispersion from each of the five frustums measured in Fig. 1b. The modal assignments are in good qualitative agreement with the analytical calculations. The color lines and symbols, along with the (lmn) designations are provided for easier comparison with the calculations. The dispersion of the three observed resonances within the far-field reflection are also provided as the open circles and correspond to the (110), (210) and (310) branches. The error bars represent a single standard deviation of the resonance peak position uncertainties determined by the least-squares fitting algorithm. Some uncertainties are smaller than the dot size.

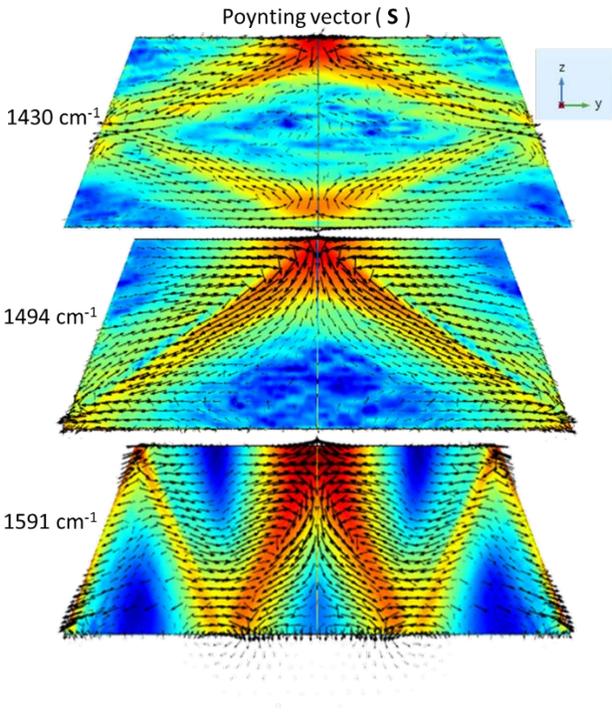

**Fig. 4**: Three dimensional electromagnetic calculations of the cross sectional electromagnetic field amplitude for a frustum with $A_r$ =2.3 illuminated by an electric point dipole located the center of the frustum, 50 nm above the top surface, at 1436 cm$^{-1}$ (top), 1491 cm$^{-1}$ (middle) and 1591 cm$^{-1}$ (bottom). The overlaid arrows designate the Poynting vector (**S**) for the polaritonic modes. Note that the direction of the field cycling and thus the predominant direction of the Poynting vector are reversed between frequencies lower (1430 cm-1) and above (1591 cm-1) the principle (110) resonance of the hBN frustum.

# Supplemental Information for

**Nanoscale mapping and spectroscopy of non-radiative hyperbolic modes in hexagonal boron nitride nanostructures**

Lisa Brown,[¥,1,2] Marcelo Davanco,[¥,1] Zhiyuan Sun,[3] Andrey Kretinin,[4] Yiguo Chen,[5,6] Joseph R. Matson,[7] Igor Vurgaftman,[8] Nicholas Sharac,[9] Alexander Giles,[8] Michael M. Fogler,[3] Takashi Taniguchi,[10] Kenji Watanabe,[10] Kostya Novoselov,[4] Stefan Maier,[5] Andrea Centrone,[1*] Joshua D. Caldwell[7,8*]

**Analytical calculation of HPhP dispersion for hBN slab**

For direct comparison to prior work, analytical calculations of the hyperbolic phonon polariton (HPhP) dispersion of a 256-nm thick slab of hexagonal boron nitride (hBN) is provided in Fig. S1. As discussed in the main text, the in-plane wavevector, $q_t$ is a continuous variable, with only the thickness dictating a quantization of the out of plane momentum, leading to the quantum number $l$.

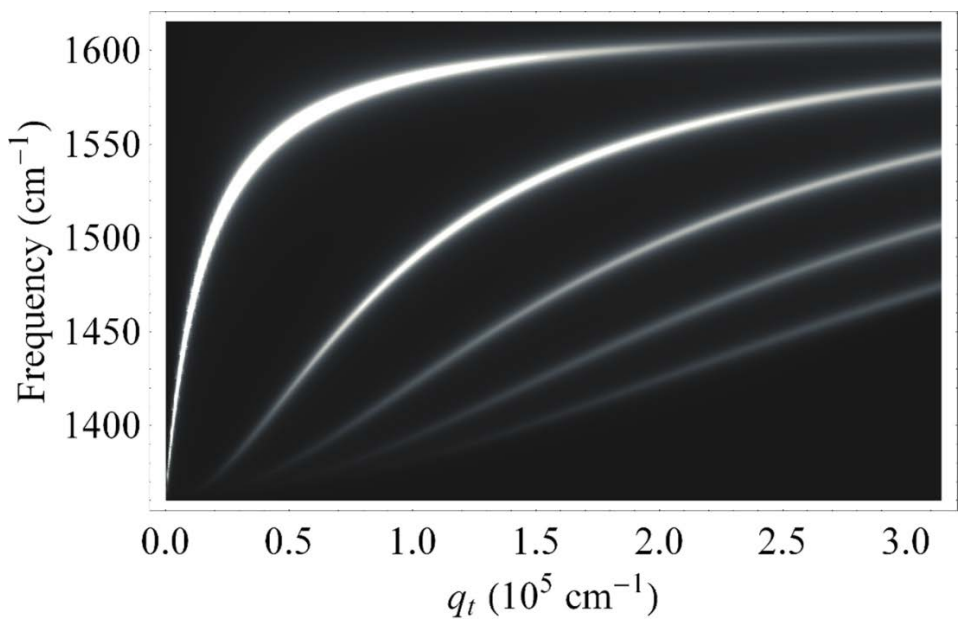

*Figure S1: **Analytical calculation of HPhP dispersion for a 256-nm thick hBN slab.** For clarity, only the first five branches ($l = 1 - 5$) are provided. Each of the five branches corresponds to an increasing value for the quantized z momentum, as described in previous work.[30] Because the hypothetical slab described here and frustum nanostructures discussed in the main text have the same thickness, these calculations highlight the additional complexity that is induced by the HPhPs confinement to three dimensional cavities where all three quantum numbers ($l, m, n$) become discrete.*

**PTIR Spectra and Analysis**

As discussed in the main text, the most interesting features in the PTIR spectra were observed using p-polarization. A direct comparison of the PTIR spectra with s- and p-polarization obtained at the frustum centers is shown in Fig S2.

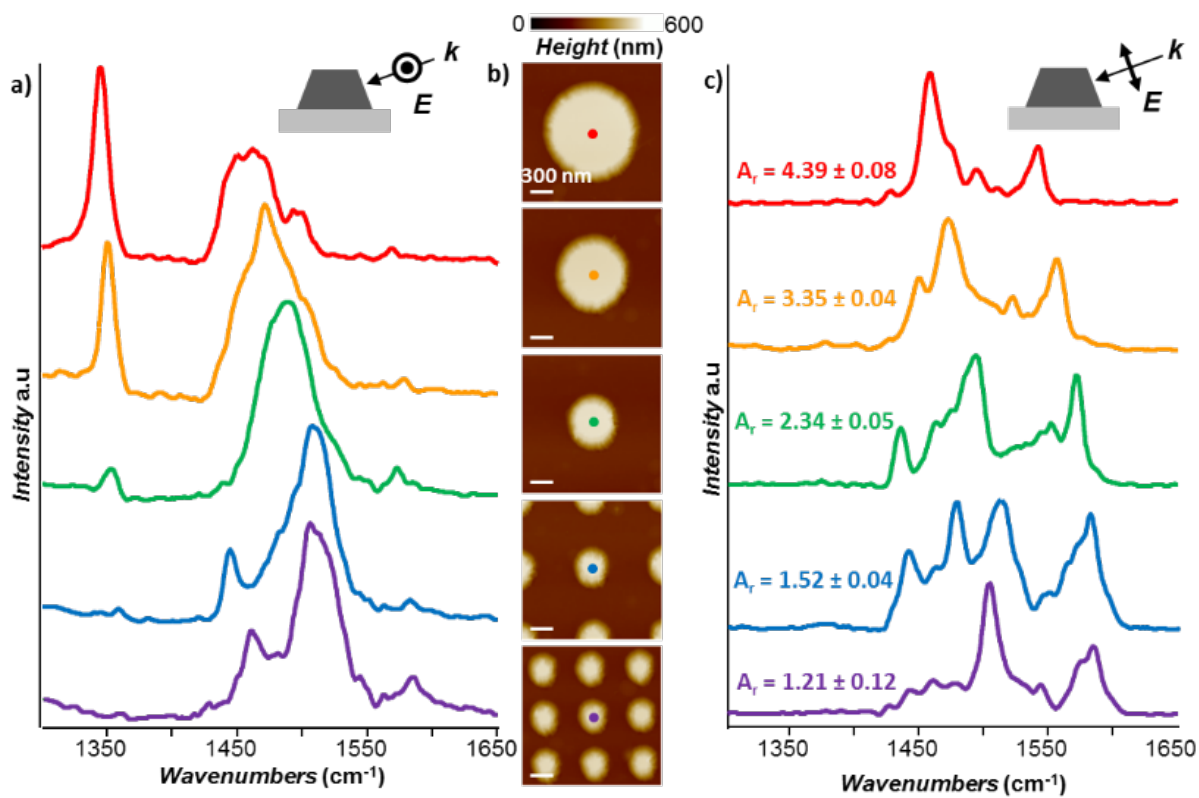

*Figure S2: s- and p-polarization PTIR spectra comparison: As discussed in the main text, the most interesting features in the PTIR spectra were observed using p-polarization. a) PTIR absorption spectra (s-polarization) obtained by positioning the AFM tip at the center of representative frustums of different aspect ratios. The pure s-polarized spectra are dominated by the fundamental (110) HPhP mode with the higher l modes observed in some spectra at lower frequencies. In the larger structures, a prominent feature is also observed at the spectral location of the TO phonon (≈1360 cm$^{-1}$). Additional weak features are observed at frequencies higher than the (110) HPhP, however, these features are much more intense and easier to identify when p-polarization is used. b) AFM topography and of five representative frustums with different aspect ratio (same as Fig 1b in the main text); scale bars are 300 nm. c) PTIR absorption spectra (p-polarization) obtained by positioning the AFM tip at the center of representative frustums of different aspect ratios (same as Fig 1d). The schematic of the incident polarization used in the experiments is provided as insets in panel a,c). The AFM tip has the effect to amplify the p-polarization component of the incident electric field because the tip polarization is much larger along the tip long axis. The spectra in panels a and c are displayed with an offset for clarity.*

The PTIR absorption maps also provide richer information when p-polarization is used as evidenced by the PTIR maps collected using s-polarized excitation (Fig. S3) display very little spatial variation in comparison to the p-polarized plots (Fig. 2c in main text). This suggests that the p-polarized component of the incident excitation field is necessary to stimulate the dark HPhPs efficiently. For s-polarization, and consistently with the PTIR absorption maps, the PTIR absorption spectra (Fig. S3e) are not very sensitive to the tip-location, again in direct contrast to the measurements with p-polarization (Fig. 1e in main text).

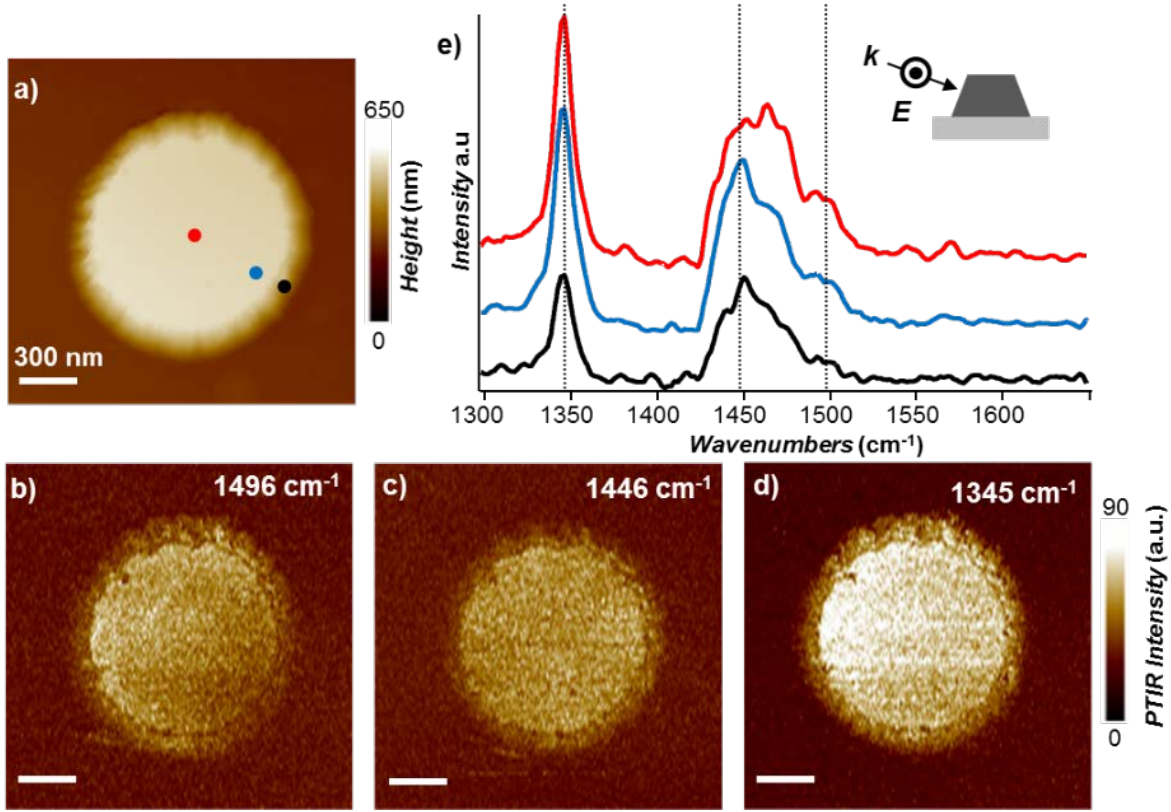

*Fig S3. s-polarization PTIR absorption maps and spectra:* *a) AFM topography for $A_r = 4.39$ frustum and corresponding PTIR maps obtained with s-polarization at b) 1496 cm$^{-1}$, c) 1446 cm$^{-1}$, d) 1496 cm$^{-1}$. e) s-polarization PTIR absorption spectra obtained at the color-coded position indicated in panel a). with s-polarization both the PTIR absorption maps and spectra display very little spatial variation (compare Fig S3 with Fig 1 and Fig S6). The spectra in panel e are displayed with an offset for clarity.*

To increase the signal to noise ratio and to better differentiate the weakest absorption features in the PTIR spectra, the PTIR spectra were smoothed over 7 adjacent points. A representative comparison between as recorded (blue) and smoothed (red) spectra is provided in Fig. S4, for the frustum with $A_r = 2.34$. Note that no apparent loss of spectral features is incurred via the smoothing function.

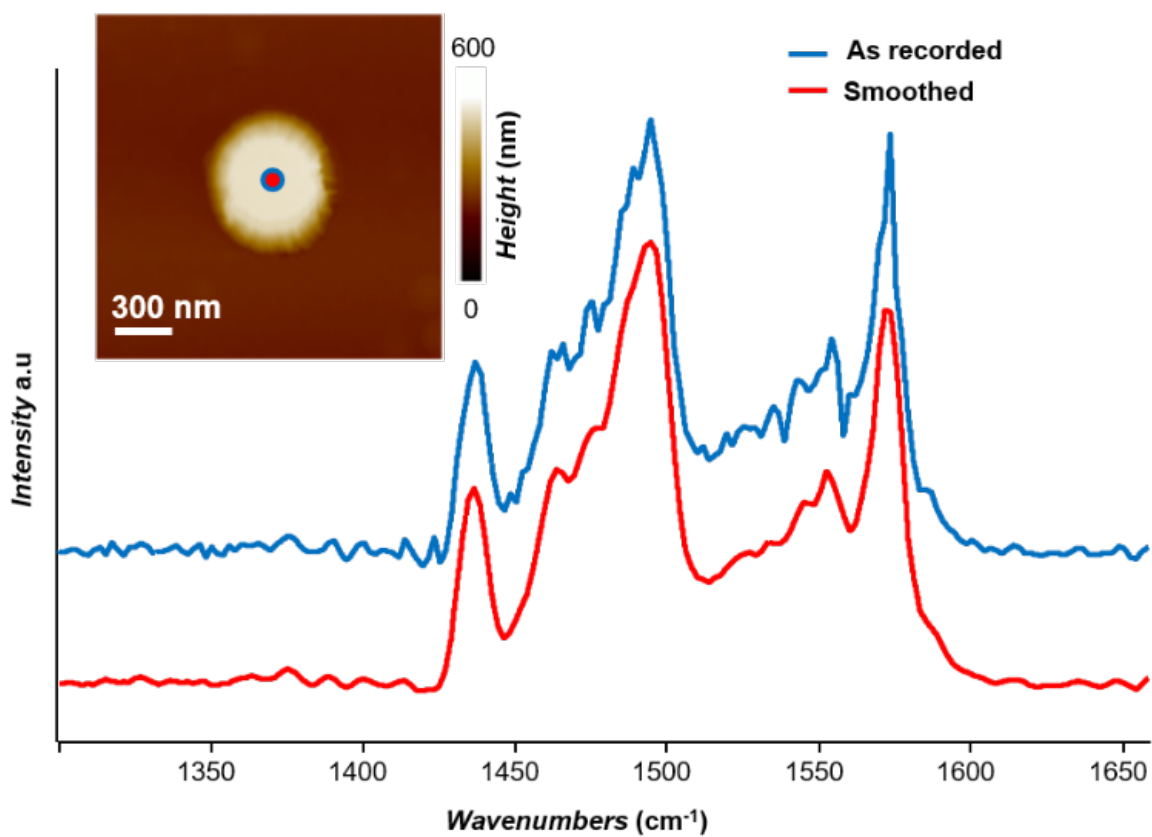

***Fig S4. Smoothing of PTIR spectra:*** *A smoothing function using the local averaging over 7 measured points was utilized to improve the signal to noise in the reported spectra. The as-recorded (blue) and smoothed spectra (red) from the center of a frustum with $A_r = 2.34$ are provided for comparison. Inset: the spatial locations of the measurements on the frustum is provided by the colored dots. The spectra are displayed with an offset for clarity.*

To compare the various HPhP modes in the PTIR spectra to the analytical calculations of HPhPs within hBN ellipsoids, least-squares fitting of the various peak positions was performed. An example of one such fit is provided in Fig. S5 for the $A_r = 4.39$ frustum. The assignment of the peaks was provided by using the analytical calculations for the ellipsoidal particles as a guide. Similarly, the analysis of the peak positions in far-field reflectance spectra were also undertaken, with only the $(l10)$ limb being detected under such conditions. The comparison of the independently determined spectral positions of the most easily identifiable modes in the far-field reflectance and PTIR spectra further support such assignments.

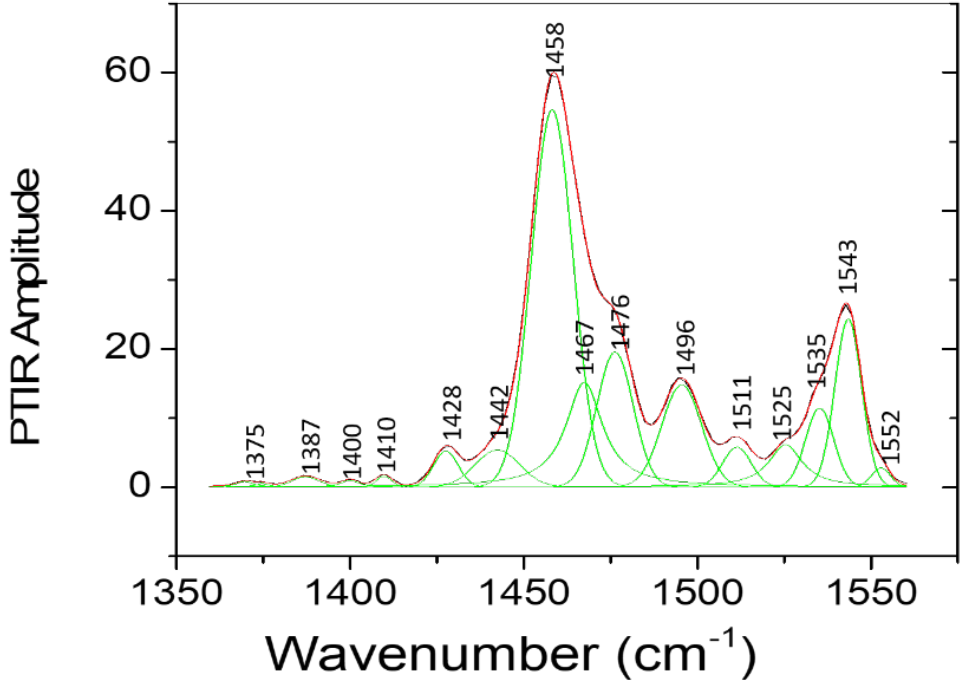

*Fig S5. Least-squares fitting of the absorption peaks in the PTIR spectra:* *Least-squares fitting of the PTIR spectra was carried out to assign the HPhP modes identified in the PTIR spectra to the corresponding modes obtained with the analytical calculations within hBN ellipsoids. The main panel provides a representative result for the fitting procedure for the $A_r = 4.39$ frustum. The assignment of the peaks was provided by using the analytical calculations for the ellipsoidal particles as a guide. Similarly, the analysis of the peak positions in far-field reflectance spectra were also undertaken (not shown), with only the $(l10)$ limb being detected under such conditions. The comparison of the independently determined spectral positions of the most easily identifiable modes in the far-field reflectance and PTIR spectra further support the assignments.*

**PTIR Absorption Maps**

In the main text the PTIR absorption images for the $A_r = 2.34$ frustum (Fig 2c), were presented, showing the spectral evolution of the polaritonic modes. However, to demonstrate that this evolution is indeed continuous and observed in all structures investigated, we present the summation of all PTIR images arranged in a matrix shown in Fig. S6. In the matrix, each column corresponds to a specified $A_r$ value (displayed scale bars vary by column) and spatial modes exhibiting similar features are aligned in the same row. From this comparison, it is quite clear that this evolution in the absorption profiles is indeed general in nature. This can be explained by the fact that the angle of HPhP propagation with respect to the z-axis is dictated by the material dielectric function, as discussed in the main text. The spectral shift between nanostructures for the similar PTIR plots is therefore dictated by the change in geometry, thought varying $A_r$. For instance, for a larger width at the same height (higher $A_r$), a larger angle is needed for the HPhP launched at the center of the top surface to extend directly to the bottom corners of the frustum.

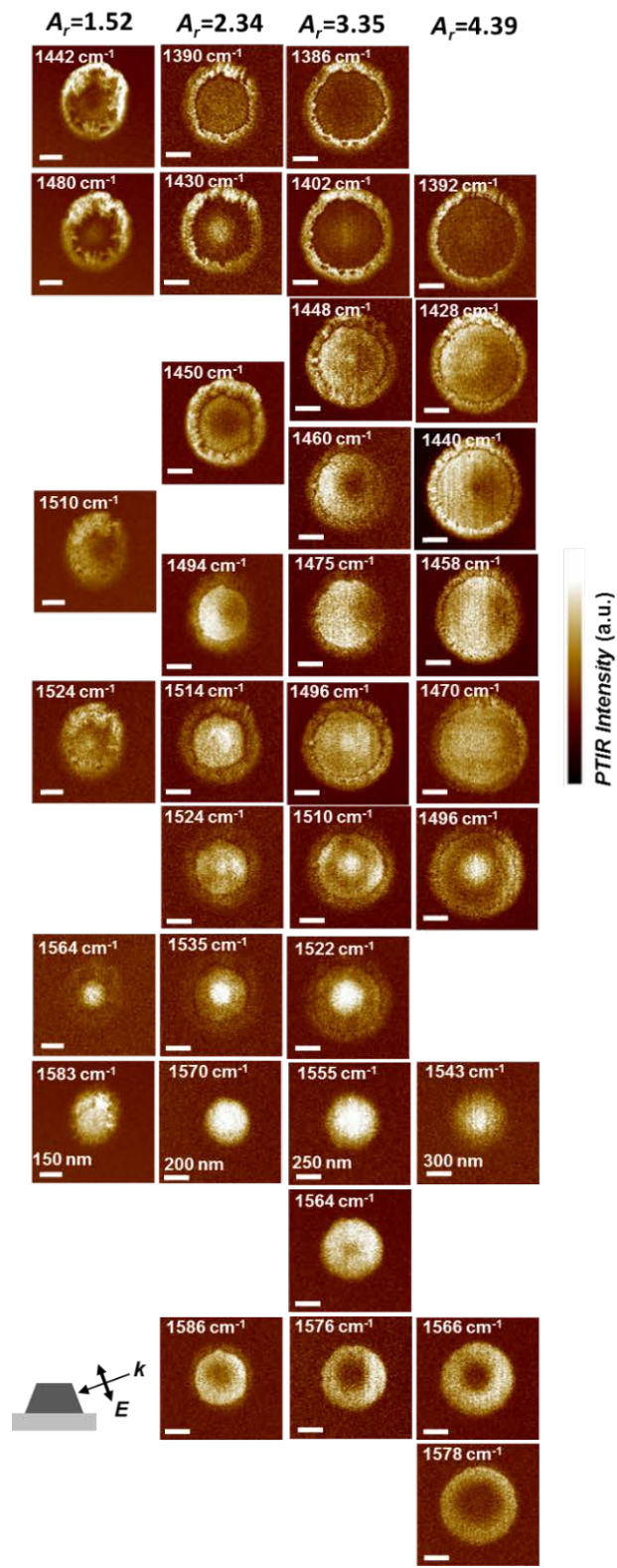

*Fig S6. PTIR maps (p-polarization) for frustums of different $A_r$:* the PTIR maps in each column corresponds to frustums with different $A_r$ (scale bars vary by column), and the maps (polaritonic modes) in the same row display similar near-field patterns suggesting that the modes are structure-specific.

This can be better appreciated by comparing the plots in Fig. S7, which reveal a clear picture of the evolution of polaritonic fields for varying $A_r$ and excitation frequency, over the nanocone cross-sections. Near-field plots with approximately the same pattern are provided in each column, with the patterns chosen to match the three originally presented in Fig. 4 of the main text. The aspect ratios chose here are close to the experimentally measured ones. Note that in order to meet the same condition within a changing $A_r$ frustum, that the angle of propagation must also change. As the angle of HPhP propagation is dictated by the frequency-dispersive material dielectric function, it is therefore a function of frequency. Thus, to realize the same near-field pattern in a larger $A_r$ frustum, the incident frequency must be reduced. It is important to note, though, that the origin of the resonances is not tied to an integer number of HPhP bounces within the structure, and thus are nominally independent from the near-field distributions.

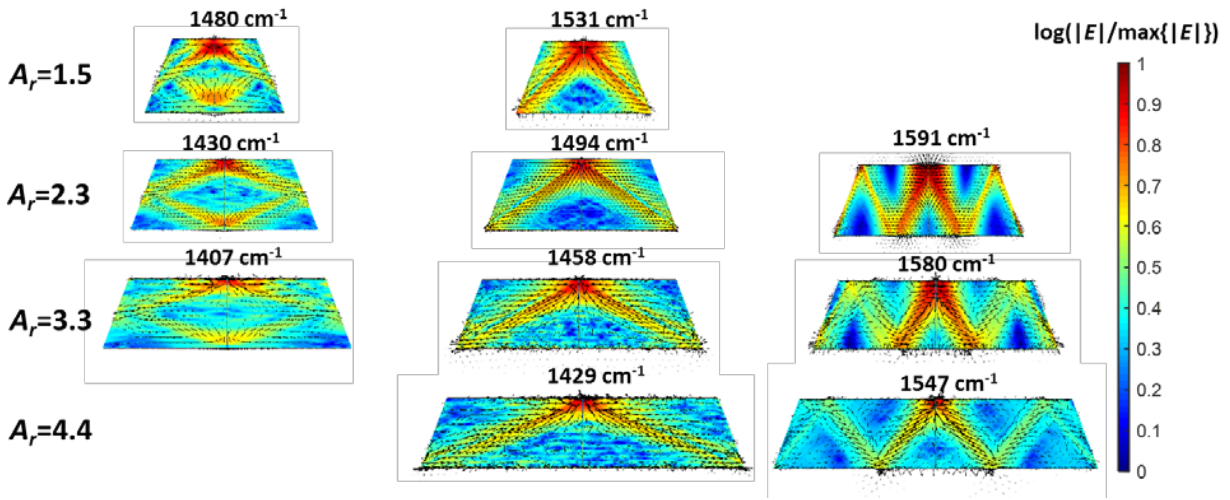

*Fig S7. Finite element calculation of the polariton electric field profile:* *Cross-sectional electric field magnitude (color scale) and Poynting vector plots (black arrows) of polariton resonances in frustums of varying aspect ratios, upon excitation with a vertical electric point dipole 50 nm above the center of top surface, at the marked frequencies. Each row corresponds to a different aspect ratio, and the resonances in the same column display similar field patterns. This shows that, despite significant variations in geometry, similar near-field patterns arise, albeit at spectrally shifted frequencies.*